\documentclass [11pt]{article}
\usepackage{amsmath,amsthm,amsfonts,amscd,eucal,latexsym,amssymb}
\usepackage{epsfig}   
\input xy
\xyoption{all}
\newsymbol\bt 1202           %%% \boxtimes 

\def\sume{{\sum\limits_{\epsilon=\pm 1}}}
\def\sumep{{\sum\limits_{\epsilon^\prime=\pm 1}}}

\def\bC{{\mathbb C}}           %%%  complex numbers and so on 
 
\def\bH{{\mathbb H}}

\def\bM{{\mathbb M}} 
 
\def\bR{{\mathbb R}}

\def\bS{{\mathbb S}}

\newsymbol \rest 1316         %%% restriction symbol 

\def\mC{{\mathcal{C}}}
\def\mD{{\mathcal{D}}}
\def\mch{{\mathcal{H}}} 
\def\mS{{\mathcal{S}}}

\def\beq{\begin{eqnarray}}
\def\eeq{\end{eqnarray}}

\newcounter{proposition}[section]
\newcounter{theorem}[section]
\newcounter{lemma}[section]
\newcounter{definition}[section]
\newcounter{remark}[section]

\def\theproposition{\thesection.\arabic{proposition}}
\def\thetheorem{\thesection.\arabic{theorem}}
\def\thelemma{\thesection.\arabic{lemma}}
\def\thedefinition{\thesection.\arabic{definition}}
\def\theremark{\thesection.\arabic{remark}}

\def\s #1 {\section{#1}}

\def\ssa #1 {\ifhmode{\par}\fi\refstepcounter{subsection}
  \noindent {\bf\thesubsection}. {\em #1}.\quad
  \addcontentsline{toc}{subsection}{\protect\numberline{\thesubsection} #1}%
  }
  
\def\sssa #1 {\ifhmode{\par}\fi\refstepcounter{subsubsection}
  \noindent {\bf\thesubsubsection}. {\em #1}.\quad
  \addcontentsline{toc}{subsubsection}{\protect\numberline{\thesubsubsection} #1}%
  }  

\def\ssb #1 {\ifhmode{\par}\fi\refstepcounter{subsection}
  \noindent {\bf\thesubsection.} {\em #1.}\quad
  \addcontentsline{toc}{subsection}{\protect\numberline{\thesubsection} #1}%
  }

\def\proposizione {\ifhmode{\par}\fi\refstepcounter{proposition}
  \noindent {\bf Proposition \theproposition}. \quad}
\def\teorema {\ifhmode{\par}\fi\refstepcounter{theorem}
  \noindent {\bf Theorem \thetheorem}. \quad}
\def\lemma {\ifhmode{\par}\fi\refstepcounter{lemma}
  \noindent {\bf Lemma \thelemma}. \quad}
\def\definizione {\ifhmode{\par}\fi\refstepcounter{definition}
  \noindent {\bf Definition \thedefinition}. \quad}
\def\remark {\ifhmode{\par}\fi\refstepcounter{remark}
  \noindent {\bf Remark \theremark}. \quad}

\begin{document} 
 
%%%%%%%%%%%%%   Title %%%%%%%%%%%%%%%%%%%%%%%%%% 
 
\par 
\LARGE 
\noindent 
{\bf Projecting Massive Scalar Fields to Null Infinity} \\
\par 
\normalsize 
 
%%%%%%%%%%%%%%%%%%%%%%%%%%%%%%%%%%%%%%%%%%%%% 
 
%%%%%%%%%%%% Authors %%%%%%%%%%%%%%%%%%%%%%%%%%% 

\noindent {\bf Claudio Dappiaggi$^{1,a}$},

\par
\small
\noindent $^1$ Dipartimento di Fisica Nucleare e Teorica, 
Universit\`a di Pavia, via A.Bassi 6 I-27100 Pavia, Italy.\smallskip

\noindent $^a$  claudio.dappiaggi@pv.infn.it,

\normalsize

\vskip .2cm

\small 
\noindent {\bf Abstract} 

It is known that, in an asymptotically flat spacetime, null infinity
cannot act as an initial-value surface for massive real scalar fields.
Exploiting tools proper of harmonic analysis on hyperboloids and global
norm estimates for the wave operator, we show that it is possible to 
circumvent such obstruction at least in Minkowski spacetime. Hence we
project norm-finite solutions of the Klein-Gordon equation of motion in
data on null infinity and, eventually, we interpret them in terms of
boundary free field theory.
\noindent 

\normalsize

\s{Introduction.}
In the study of  classical fields over four dimensional Lorentzian 
curved backgrounds, Penrose conformal completion techniques have played 
since their introduction a pivotal role. 

In particular the related notion of asymptotic simplicity/flatness 
entails the embedding of a (four dimensional) physical spacetime 
$(M^4, g_{\mu\nu})$ as a bounded open set in an unphysical background 
$(\widehat M^4, \widehat g_{\mu\nu})$ being $\widehat g$ a conformal 
rescaling of $g$. In this setting the image of $M^4$ in $\widehat M^4$
can be naturally endowed with a boundary structure usually referred to 
as $\Im^\pm$ {\it i.e.} future or past null infinity.
 
Heuristically the endpoint of all the null geodesics in 
$(M^4, g_{\mu\nu})$, $\Im^\pm$ is thus the geometrical locus where the 
trajectory of  zero rest mass particles end. Hence it is manifest how 
null conformal boundaries 
can be exploited as a powerful tool to study either the asymptotic 
properties of radiation fields associated to massless wave equations, 
either the scattering properties of massless fields \cite{Mason}.

Furthermore, from the perspective of quantum field theory over curved 
backgrounds, $\Im^\pm$ plays a key role in the realization of the 
holographic principle. The latter conjectures that the information of any 
field theory on a D-dimensional Lorentzian background $M$ can be 
recovered by means of a suitable second field theory constructed
over a codimension one submanifold $\Sigma$ embedded in  $M$. Hence, in
asymptotically flat spacetimes it is natural to conjecture that the role
of $\Sigma$ is played by the null conformal boundary and this idea has
been successfully investigated both at a classical and at a quantum level 
in \cite{Dappiaggi}.

To better understand the main rationale underlying the success 
of Penrose conformal techniques from a field theoretical perspective, let
us consider a working example, namely the massless Klein-Gordon real
scalar field $\psi$ conformally coupled to gravity in a globally
hyperbolic and asymptotically flat spacetime $M^4$. Barring a few 
technical assumptions, each  solution of $\left[\square_{g}-\frac{R}{6}
\right]\psi=0$ with compactly supported initial data on a Cauchy surface
can be mapped into a solution of
\beq\label{prima}
\widehat\square_{\widehat g}\widetilde\psi-\frac{\widehat{R}}{6}
\widetilde\psi=0,
\eeq
where $\widetilde\psi\doteq \Omega^{-1}\psi$. 
Although $\widetilde\psi$ is strictly defined only over the image of
$M^4$ in $\widehat{M}^4$, global hyperbolicity of $\widehat{M}^4$ and
uniqueness of solutions for second order hyperbolic PDE, allows us to
extend $\widetilde\psi$ to a smooth solution for \eqref{prima} over all
$\widehat{M}^4$. Accordingly we can define the projection of $\widetilde\psi$
over the boundary $\Im^\pm$ simply as its restriction: 
$\Psi_{\pm}\doteq\widetilde\psi|_{\Im^\pm}\in C^\infty(\Im^\pm)$. It is
$\Psi$ the key ingredient to study properties of bulk physical phenomena
starting from boundary data in the unphysical spacetime as exploited, to
quote just a few examples, in \cite{Dappiaggi, Mason}.

Nonetheless the situation is not heavenly as it may seem since the above
construction drastically fails whenever one considers massive
fields. Even in the simplest situation of the Klein-Gordon scalar field
on flat Minkowski spacetime, conformal invariance of the equation of
motion is broken. Furthermore it has been argued in
\cite{Helfer, Winicour} that $\Im^\pm$ cannot be used as an initial value
surface for massive fields and that it is not possible to project any
solution of $\left[\square_g-m^2\right]\psi=0$ into a smooth function
over $\Im^\pm$. This result has been established with an
elegant argument in \cite{Helfer}: the space of sections of any vector 
bundle on $\Im^\pm$ which is homogeneous for the action of the Poincar\'e 
group carries only massless representations\footnote{A reader familiar
with Penrose compactification techniques could argue that the relevant
symmetry group on null infinity is not the Poincar\'e but the BMS group
which is the semidirect product between the proper ortochronous component
of the Lorentz group and the smooth functions over the 2-sphere thought
as an Abelian group under addition. Nonetheless, since in Helfer
construction, the key role is played by the translational subgroup of
the Poincar\`e group, the result can be extended also in a BMS framework
remembering that it exists a four dimensional normal subgroup of the
full BMS group homeomorphic to $T^4$ \cite{Dappiaggi}.}.   

Hence it seems impossible to exploit the powerful means of Penrose
compactification whenever we deal with solutions of partial differential
equations containing a term proportional to a scale length such 
as the mass. In other words, since the information of the data evolving 
to infinity along causal timelike curves flows in the unphysical 
spacetime $\widehat M^4$ to future timelike infinity $i^+$  (a 
codimension 2 submanifold of $\widehat M^4$ hence not a proper boundary), it
seems impossible to exploit null infinity as a tool to study massive
fields.

The aim of this paper is to provide a way to circumvent the
above obstruction at least in Minkowski background. In particular we 
will exploit both tools of harmonic analysis and global norm estimates
for the wave equation in order to project a solution for the massive
Klein-Gordon equation of motion into meaningful data over null infinity. 

More in detail, the outline of the analysis and hence of the paper will
be the following: in the next subsection we recollect some basic details
about the notion of asymptotic flatness. In section 2, instead, we
specialise to Minkowski background and we consider solutions of the 
massive Klein-Gordon equation of motion satisfying a finite norm 
condition in such a way that their Fourier transform is a square 
integrable function over the mass hyperboloid $\bH_m$. 
Exploiting a few results due to Strichartz on harmonic
analysis over hyperboloids we shall introduce a unitary map between two
copies of $L^2(\bH_m)$ and the space of square integrable function over 
the light cone $C$. Furthermore such a map will also act as an 
intertwiner between the quasi-regular representations of the Lorentz 
group on $L^2(\bH_m)$ and $L^2(\mC)$.

Afterward, as a first step, we exploit global norm estimates to associate 
to each square integrable function over the light cone a norm finite 
solution for the wave equation in Minkowski spacetime. By means of 
Penrose compactification techniques and trace theorems, we project these 
functions on null infinity.

Eventually, in section 3, we show how the projected data can be
interpreted in terms of a diffeomorphism invariant field theory
intrinsically constructed over null infinity.\\

\ssa{On asymptotically flat spacetimes}\label{AFS}
In this section we recollect some known facts about the definition and
the properties of asymptotically flat spacetime. Although we are going
to work in Minkowski background, the following summary can be useful for
a twofold reason: from one side in section 3 we shall interpret the
projection of the data from a bulk massive scalar field in terms of a
field theory on future null infinity whereas, from the other side, we
look at this paper as the first step to solve the same problem on a
generic asymptotically flat spacetime. Hence it could be interesting to
understand where our construction relies on properties specific of
Minkowski spacetime and where, on the opposite, out results could be
traded to a more general scenario.

In the literature there are several different notions of asymptotic
flatness at (future or past) null infinity which are obviously all 
equivalent if the bulk spacetime is Minkowski; hence a reader familiar
to any of these can skip to next section without a second thought. We 
shall instead adopt the specific definition first introduced
by Friedrich (see \cite{Friedrich} and references therein from the same 
author) of a class of spacetimes which are flat at future 
null infinity and they admit future time completion at $i^+$. The reason
for this choice lies in the realm of quantum field theory of curved
background. In particular in \cite{Dappiaggi} it has been shown that
it is possible to project the Weyl *-algebra of observables for a real
massless scalar field in Minkowski spacetime as a subsector of a
suitable counterpart at null infinity because the Lichnerowicz
propagator for the wave operator is strictly supported on the light
cone. On the opposite, in a generic curved background, a priori this 
does not held true since the support includes a tail strictly contained 
in the cone and, hence, in the conformal completion language propagating 
at future timelike infinity. Thus in order to recast the result of
\cite{Dappiaggi} in a generic scenario Friedrich definition is the most
appealing (to this avail see the analysis in \cite{Moretti}).

In detail a four dimensional future time oriented spacetime $M^4$ with a
smooth metric $g_{\mu\nu}$ which solves the vacuum Einstein equation is 
called an \emph{asymptotically flat spacetime with future time infinity
$i^+$} if it exists a second four dimensional spacetime $(\widehat M ,
\widehat{g}_{\mu\nu})$ with a preferred point $i^+$, a diffeomorphism
$\lambda: M \to\lambda(M)\subset\widehat M$ and a non negative scalar
function $\Omega$ on $\lambda(M)$ such that
$\widehat{g}=\Omega^2\lambda^*g$ and the following facts hold:
\begin{enumerate}
\item $J^-(i^+; \widehat{M})$ is closed and
$\lambda(M)=J^-(i^+)\setminus\partial J^-(i^+; \widehat{M})$. Moreover 
$\partial \lambda(M) = \Im^+\cup i^+$ where $\Im^+\doteq \partial J^-(
i^+;\widehat{M})\setminus\left\{i^+\right\}$ is future null infinity.
\item $\lambda(M)$ is strongly causal.
\item $\Omega$  can be extended to a smooth function on $\widehat{M}$.
\item $\left.\Omega\right|_{\partial J^-(i^+;\widehat{M})}= 0$, but
$d\Omega(x)=0$ for $x\in\Im^+$ and $d\Omega(i^+)=0$, but $\widehat\nabla
_\mu\widehat\nabla_\nu\Omega(i^+)=-2\widehat{g}_{\mu\nu}(i+)$.
\item If $n^\mu\doteq\widehat{g}^{\mu\nu}\widehat\nabla_\nu\Omega$ then
it exists a strictly positive smooth function $\omega$, defined in a 
neighbourhood of $\Im^+$ and satisfying $\widehat\nabla_\mu(\omega^4 
n^\mu)=0$ on $\Im^+$, such that the integral curves of $\omega^{-1}n^\mu$
are complete on $\Im^+$.
\end{enumerate}

From now we shall refer to $\lambda(M)$ simply as $M$ since no confusion
will arise in the manuscript due to this identification. Furthermore we
point out that, with minor adaption, the above definition can be recast
for spacetimes which are asymptotically flat with past time infinity
$i_-$ and henceforth we shall refer only to $\Im^+$ though the reader is
warned that all our results hold identically for $\Im^-$.

Thus let us consider any asymptotically flat spacetime as per the
previous definition; the metric structure of future null infinity is not 
uniquely determined but it is affected by a gauge freedom in the choice 
of the compactification factor namely, if we rescaled $\Omega$ as 
$\omega\Omega$ with $\omega\in C^\infty(\Im^+,\bR^+)$, the topology and 
the differentiable structure of future null infinity is left unchanged. 
Hence the difference between the 
possible geometries for the conformal boundary is caught by equivalence 
classes of the following triplet of data $(\Im^+,n_a, h^{ab})$ where 
$\Im^+$ stands for the $S^2\times\bR$ topology of null infinity,
$n_a\doteq\widehat\nabla^a\Omega$ (being $\widehat\nabla$ the covariant 
derivative with respect to $\widehat g_{ab}$) and $h_{ab}\doteq\widehat 
g_{ab}|_{\Im^+}$. Two triplets $(\Im^+,n_a, h^{ab})$ and $(\Im^+,n^
\prime_a, h^{\prime ab})$ are called equivalent iff it exists a gauge 
factor  $\omega$ such that $h^\prime_{ab}=\omega^2 h_{ab}$ whereas 
$n^{\prime a}=\omega^{-1}n^a$.

The set of all these equivalence classes is {\it universal} in the sense
that, given any two asymptotically flat spacetime $M_1$ and $M_2$ with
associated triplets $(\Im^+_1,n_{1 a}, h^{ab}_1)$ and $(\Im^+_2,n_{2 a}, 
h^{ab}_2)$, it always exists a diffeomorphism $\gamma\in
Diff(\Im^+_1,\Im^+_2)$ such that $\gamma^*h^{ab}_2=h^{ab}_1$ and
$\gamma_*n_{1 a}=n_{2 a}$.

The set of all group elements $\gamma\in Diff(\Im^+,\Im^+)$ mapping a
triplet into a gauge equivalent one\footnote{Although at first sight we
are considering a subgroup of the whole set of diffeomorphism, one
should take into account that the constraint we impose is equivalent to
require that the bulk geometry is left unchanged {\it i.e.} we are
working on a fixed background.} is called the Bondi-Metzner-Sachs
group (BMS). It is always possible to choose $\omega$ in such a way that
on null infinity we can introduce the so-called Bondi frame 
$(u,z,\bar{z})$ where $u$ is the affine parameter along the null
complete geodesics generating $\Im^+$ and $(z,\bar{z})$ are the complex
coordinates construct out of a stereographic projection from
$(\theta,\varphi)\in\bS^2$, then the BMS group is $SO(3,1)\ltimes
C^\infty(\bS^2)$ acting as
\begin{gather}\label{BMS1}
u\longrightarrow
u^\prime=K_\Lambda(z,\bar{z})\left(u+\alpha(z,\bar{z})\right),\\
z\longrightarrow\Lambda z\doteq\frac{az+b}{cz+d},\quad
\bar{z}\longrightarrow\Lambda \bar{z}\doteq\frac{\bar{a}\bar{z}+\bar{b}}
{\bar{c}\bar{z}+\bar{d}},\label{BMS2}
\end{gather} 
where $\Lambda$ is identified with the matrix $\left[
\begin{array}{cc}
a & b\\
c & d
\end{array}\right]$ whereas
$$K_\Lambda(z,\bar{z})=\frac{1+|z|^2}{|az+b|^2+|cz+d|^2}.$$
A direct inspection of this formula shows that the BMS group is a
regular semidirect product and it is much larger than the Poincar\'e 
group. In a generic scenario such a problem cannot be easily overcome 
though one can recognise that any element in the Abelian ideal 
$C^\infty(\bS^2)$ can be expanded in real spherical harmonics as 
$$\alpha(z,\bar{z})=\sum\limits_{l=0}^1\sum\limits_{m=-l}^l\alpha_{lm}
S_{lm}(z,\bar{z})+\sum\limits_{l=2}^\infty\sum\limits_{m=-l}^l\alpha_{lm}
S_{lm}(z,\bar{z}).\quad\forall\alpha(z,\bar{z})\in C^\infty(\bS^2)$$
Here we have separated the set of first four components - known as the
translational component of the BMS group - since it is homeomorphic to
the Abelian group $T^4$. Furthermore the following proposition holds:\\

\proposizione\label{transl}{\em The subset $SO(3,1)\ltimes T^4$ of the 
BMS group made of elements $\left(\Lambda,\alpha(z,\bar{z})\right)$, 
where  $\alpha(z,\bar{z})$ is a real linear combinations of the first 
four spherical harmonics, is a BMS subgroup and if we associate to 
$\alpha(z,\bar{z})$ the vector
$$a^\mu=-\sqrt{\frac{3}{4\pi}}\left(\frac{a_{00}}{\sqrt{3}},a_{1-1},
a_{10},a_{11}\right),$$
the action of $\Lambda\in SO(3,1)$ on $a^\mu$ is equivalent to the
transformation of the 4-vector in Minkowski background under the
standard Lorentz action.} \\

\noindent The proof of this theorem has been given in propositions 3.11 
and 3.12 in \cite{Dappiaggi}.\\

On the opposite we wish to underline that, in a generic asymptotically
flat spacetime, we cannot exploit this last statement to select a
preferred Poincar\'e subgroup in the BMS since, acting per conjugation
over the above $SO(3,1)\ltimes T^4$ subset with any element
$(I,S_{lm}(z,\bar{z}))\in SO(3,1)\ltimes C^\infty(\bS^2)$ with $l>1$ we
end up with a different though equivalent Poincar\'e subgroup.
Nonetheless, since in this paper we are taking into account only
Minkowski background, we can exploit a result due to Geroch, Ashtekar and
Xanthopoulos \cite{Ashtekar, Geroch} namely\\

\proposizione{\em In any asymptotically flat spacetime $(M,g_{\mu\nu})$ 
it holds
\begin{enumerate}
\item[a)] any Killing vector $\xi$ in M smoothly extends to a Killing
vector $\widehat\xi$ in $\widehat M$ and the restriction $\tilde\xi$ of 
the  latter to $\Im$ is tangent to null infinity, it is uniquely 
determined by $\xi$ and it generates a one-parameter subgroup of the BMS.
\item[b)] the map $\xi\to\tilde\xi$ is injective and, if the
one-parameter subgroup of the BMS generated by $\tilde\xi$ lies in
$C^\infty(\bS^2)$ then it must also be a subgroup of $T^4$.
\end{enumerate}}

According to the last proposition, in a Minkowski background, the
Poincar\'e isometries identify a preferred subgroup of the BMS group
{\it i.e.} the set
\beq\label{subgroup}
\mathcal{R}=\left\{(\Lambda,\alpha(z,\bar{z}))\;|\;\alpha(z,\bar{z})=a^0+
a^1\frac{z+\bar{z}} {1+|z|^2}+a^2\frac{z-\bar{z}}{1+|z|^2}+a^3
\frac{|z|^2-1}{1+|z|^2}\right\},
\eeq
which is homomorphic to $SO(3,1)\ltimes T^4$.

\section{From Massive to Massless Scalar fields on Minkowski spacetime}
\label{Massive} 

Let us consider four dimensional flat Minkowski spacetime
$\left(M^4,\eta_{\mu\nu}\right)$ and a real scalar field
$\phi:M^4\to\mathbb{R}$ satisfying the Klein-Gordon equation with
squared mass $m^2>0$:
\beq\label{KG}
\square_\eta\phi-m^2\phi=0.
\eeq
In the most general framework we should seek for tempered distributions
solutions to such PDE and their Fourier transform is a function supported 
on the mass hyperboloids $\mathbb{H}_m$ (see section IX.9 of \cite{Reed}) 
$\eta^{\mu\nu}p_\mu p_\nu=p_0^2-\sum\limits_{i=1}^3p^2_i=m^2$, being 
$p_\mu=\left(p_0, p_i\right)$ with $i=1,..,3$ the standard global 
coordinates\footnote{The
symbols are here adopted with respect to the standard high energy
physics terminology though we do not seek at the moment any physical 
interpretation of the forthcoming analysis leaving it for the
conclusions.} on each fibre of the cotangent bundle
$T^*M^4$ canonically identified as $\mathbb{R}^4\times\mathbb{R}^4$. 

The mass hyperboloids can be parameterised with the coordinates 
$r\doteq|\vec{p}\;|\equiv\left(\sum\limits_{i=1}^3p^2_i\right)^{\frac{1}{2}}
\in[0,\infty)$, $\vec{\zeta}\doteq\frac{\vec{p}}{|\vec{p}\;|}
\in\mathbb{S}^2\hookrightarrow\bR^3$ and $\epsilon\doteq\frac{p_0}{|p_0|}
=\pm 1$. The variable $\epsilon$ provides a way to distinguish in $\bR^4$ 
between the upper and lower hyperboloid and we will keep track of it for 
the sake of generality. An interested reader can adapt the following 
constructions to a single hyperboloid with minor efforts. 

Furthermore, identifying $\bH_m$ with the coset
$\frac{O(3,1)}{O(3)}$, we can endow it with the $O(3,1)$ invariant
measure $d\mu(\bH_m)=\frac{r^2}{\sqrt{r^2+m^2}}drd\zeta$. Hence we can 
take into account only the solutions of \eqref{KG} that are finite with 
respect to a suitable norm
{\it i.e.}, following the conventions of \cite{Strichartz}, it must 
exists a real number $\alpha\geq 0$ and a function $f(r,\zeta,\epsilon)$ 
such that, being $\vec{x}$ the spatial component of $x^\mu$ and $\cdot$
the standard Euclidean scalar product on $\bR^3$,
$$\phi(x^\mu)=\sume\int\limits_{\bH_m} d\mu(\bH_m)\;e^{ir
\vec{x}\cdot\vec{\zeta}}e^{-i\sqrt{r^2+m^2}t\epsilon}f(r,\zeta,\epsilon)
$$
and
\beq\label{norm}
||\phi||^2_\alpha=\sum\limits_{\epsilon=\pm
1}\int\limits_{S^2}d\zeta\int\limits_0^\infty
dr\left|(r^2+m^2)^{\frac{\alpha}{2}}f(r,\zeta,\epsilon)\right|^2
d\mu(\bH_m)<\infty.
\eeq
Dropping from now any on all references to $d\mu(\bH_m)$, we shall call
the Hilbert spaces of functions satisfying \eqref{norm} as 
$L^2_\alpha(\bH_m)$ and, out of a direct inspection of the above
formula, the following chain of inclusions holds: $L^2(\bH_m)\equiv L^2_0
(\bH_m)\subset L^2_\alpha(\bH_m)\subset L^2_{\alpha^\prime}(\bH_m)$ for 
all $0<\alpha<\alpha^\prime$.

To summarise the key point, the constraint \eqref{norm} allows us a way 
to select only those solutions $\phi$ of \eqref{KG} whose
Fourier transform $f$ is at least square integrable on the mass 
hyperboloid with respect to the $O(3,1)$-invariant measure. Furthermore 
we can require the $O(3,1)$ group to act on $f$ with the quasi-regular scalar 
representation {\it i.e.} for any $\Lambda\in O(3,1)$ and for any 
$p_\mu\in\bH_m\hookrightarrow\bR^4$
\beq\label{quasiregular}
U(\Lambda)f(p_\mu)=f(\Lambda^{-1}p_\mu),\quad f\in L^2\left(\bH_m,\right)
\eeq
being $U$ unitary strongly continuous but not irreducible.

Henceforth our plan is to discuss and later to exploit the following
Strichartz result: it is possible to construct an operator 
$T$ from $L^2\left(\bH_m\right)\oplus L^2\left(\bH_m\right)$ into the 
space of square-integrable functions over the light cone with respect to
the $O(3,1)$-invariant measure and $T$ is also a unitary 
intertwiner\footnote{We recall that, given a group $G$ with the 
representations $U$ and $U^\prime$ on the Hilbert spaces $\mch$ and 
$\mch^\prime$, a bounded linear map $T:\mch\to\mch^\prime$ is called an 
intertwiner if $U^\prime(g)T=TU(g)$ for all $g\in G$.} between the 
quasi-regular $O(3,1)$-representations.\\

\ssa{From hyperboloids to light cones}\label{FHLC}

The analysis and the statements in this section are based upon the
theorems proved in \cite{Strichartz2} even though part of the
results have been independently developed also in \cite{Limic} and, by
means of integral transforms associated to horospheres.
The proof of most of the following results strongly relies upon the
embedding of the mass hyperboloid and of the light cone in $\bR^4$. All
the analysis can be recast in terms of the intrinsic structures over
these symmetric space and we refer to \cite{Rossmann} for an interested
reader.

As a starting point we shall briefly discuss and characterise some
properties of square integrable functions over the light cone. Let us 
quickly recall that the latter is the geometric locus $\mathcal{C}=\left
\{p_\mu=(p_0,p_i),\;\eta^{\mu\nu}p_\mu p_\nu=p^2_0-|\vec{p}\;|^2=0\right
\}\setminus(0,0)$ where $p_\mu=(p_0,\vec{p})=(p_0, p_i)$ with 
$i=1,...,3$ are the same global coordinates introduced in the previous 
section. As for $\bH_m$ we can set a more convenient coordinate system 
and an $O(3,1)$-invariant measure which are basically constructed with 
a limiting procedure ({\it i.e.} $m\to 0$) from  the counterpart on the 
mass hyperboloid. Namely, if we refer to $r\doteq|\vec{p}\;|\in \left(0,
\infty\right)$, $\vec{\zeta}\doteq\frac{\vec{p}}{|\vec{p}\;|}\in \mathbb
{S}^2\hookrightarrow\bR^3$ and $\epsilon\doteq\frac{p_0}{|p_0|}=\pm 1$,
the measure is $d\mu(\mC)=rdrd\zeta$. Here the two values of $\epsilon$ 
allow us to distinguish between the future and the past light cone and, 
as for the massive case, we keep track of them for the sake of 
completeness.

The next step consists of a specific characterisation for square
integrable functions over the light cone with respect to $d\mu(\mC)$.
Let us consider the set $\mD^0_\sigma$ and $\mD^1_\sigma$ respectively as 
even and odd smooth functions over $\mC$ homogeneous of degree $\sigma$ 
in the $r$-variable {\it i.e.} of the form $r^\sigma g(\zeta,\epsilon)$. 
Then the following proposition holds\\

\proposizione{\label{prop1}\em
If $\sigma=-1+i\rho$ with $\rho\in\bR$, then $\mD^0_\sigma$ and 
$\mD^1_\sigma$ can be closed to Hilbert space $\mch^0_\sigma$ and
$\mch^1_\sigma$ with respect to the norm
$$||r^\sigma g(\zeta,\epsilon)||^2_\sigma=\sume\int\limits_{\bS^2}d
\zeta|g(\zeta,\epsilon)|^2.$$
Furthermore 
\begin{enumerate}
\item the quasi-regular $O(3,1)$ scalar representation acting on
the functions over $\mC$ as
$$U^\prime(\Lambda)F(p_\mu)=F(\Lambda^{-1}p_\mu),\quad\forall p_\mu
\in\mC\hookrightarrow \mathbb{R}^4\;\wedge\;\forall F\in L^2(\mC),$$
is strongly continuous unitary and irreducible on both $\mch^0_\sigma$
and $\mch^1_\sigma$.
\item for any $F\in L^2(\mC)$ it exists a unique function $\varphi_0$ in 
$\mch^0_\sigma$ and $\varphi_1$ in $\mch^1_\sigma$ such that, calling
$F_0(p_\mu)=\frac{1}{2}\left(F(p_\mu)+F(-p_\mu)\right)$ and $F_1(p_\mu)=
\frac{1}{2}\left(F(p_\mu)-F(-p_\mu)\right)$, then 
\beq\label{int1}
||F_j(p_\mu)||_{L^2(\mC)}=\sume\int\limits_{-\infty}^\infty \frac{d\rho}
{2\pi}\int\limits_{\bS^2}d\zeta\left|\varphi_j(\rho,\zeta,\epsilon)
\right|^2,\quad j=0,1
\eeq
and 
\beq\label{int2}
F_j(r,\zeta,\epsilon)=\sume\int\limits_{-\infty}^\infty \frac{d\rho}
{2\pi}r^{-1+i\rho}\varphi_j(\rho,\zeta,\epsilon).\quad j=0,1
\eeq
\end{enumerate}
The image of the map $F\to\left(\varphi_0,\varphi_1\right)$ is onto all
pairs with a finite right hand side in \eqref{int1}.}\\

\begin{proof}
We here sketch the main details of the proof as in \cite{Strichartz2}. To
start, let us notice that the norm over $\mD^0_\sigma$ and $\mD^1_\sigma$
is well defined since, up to the sum over $\epsilon$, it is equivalent to
the norm over $L^2(\bS^2,d^2x)$ being $d^2x$ the Lesbegue measure on
$\bS^2$.

The unitarity and strong continuity of the quasi-regular representation arises due to the
$O(3,1)$-invariance of the measure on the light cone. Hence for any
$F(p_\mu)\in L^2(\mC)$ with $p_\mu\in\bR^4$ satisfying
$\eta^{\mu\nu}p_\mu p_\nu=0$, it holds:
$$\int\limits_\mC d\mu(\mC)|U^\prime(\Lambda)F(p_\mu)|^2=
\int\limits_\mC d\mu(\mC)|F(\Lambda^{-1}p_\mu)|^2=\int\limits_\mC
d\mu(\Lambda\mC)|F(p_\mu)|^2=\int\limits_\mC d\mu(\mC)|F(p_\mu)|^2,$$
where, in the second equality, we performed the coordinate change
$p_\mu\to\Lambda p_\mu$.

To prove irreducibility let us note that any function $f\in\mch^j_\sigma$
with $j=0,1$ can be decomposed in spherical harmonics {\it i.e.} 
$f(r,\zeta,\epsilon)=r^\sigma
g(\zeta,\epsilon)=\sum\limits_{l=0}^\infty\sum\limits_{m=-l}^l
a_{lm}Y_{lm}(\zeta)\epsilon^k r^\sigma$ where $k=0,1$ and the coefficients $a_{jm}$
must vanish if $j=0$ and $l+k$ is odd or if $j=1$ and $l+k$ is even.
Consider now, as a special case, a function in $\mch^j_\sigma$ with all
but one of the coefficients $a_{lm}$ equal to zero. We show now that the
action of the quasi-regular $O(3,1)$ representation generates a second
function with the coefficients $a_{l+ 1, m}\neq 0$. To this avail let
us choose an element of $SU(1,1)\subset O(3,1)$ parameterised by an
angle $\alpha$, apply it to $f$ and then let us differentiate with 
respect to $\alpha$. The resulting function $f^\prime$ evaluated in 
$\alpha=0$ is 
$$f^\prime(r,\zeta,\epsilon)=\frac{(\sigma-l)(l+1)}{{1+2l}}r^\sigma Y_{l
+1}(\zeta)\epsilon^{k+1}.$$  
Since all these operations should map any irreducible subspace of
$\mch^j_\sigma$ into itself, the statement in point 1. of the theorem 
holds.

To demonstrate point 2. let us associate to any  $F(p_\mu)\equiv
F(r,\zeta,\epsilon)\in L^2(\mC)$, the functions
$g_j(r,\zeta,\epsilon)=rF_j(r,\zeta,\epsilon)$ with $j=0,1$. Hence for
each $j$
$$\int\limits_\mC
d\mu(\mC)|F_j(r,\zeta,\epsilon)|^2=\int\limits_{\bS^2}d\zeta\int\limits_0^\infty
dr\; r^{-1}|g_j(r,\zeta,\epsilon)|^2<\infty,$$
which implies that $\int\limits_0^\infty dr r^{-1}|g_j(r,\zeta,\epsilon)|
^2<\infty$ per Fubini's theorem. Hence we can apply Mellin inversion
theorem to write $g_j(r,\zeta,\epsilon)=\int\limits_{-\infty}^\infty
d\rho\; r^{i\rho}\varphi_j(\rho,\zeta,\epsilon)=\int\limits_{-\infty}^\infty
d\rho\; e^{i\rho\ln(r)}\varphi_j(\rho,\zeta,\epsilon)$. The last identity
suggests us to apply Plancherel theorem to conclude that
$\int\limits_0^\infty\;d\ln(r)|g_j(r,\zeta,\epsilon)|^2=\int\limits_{-
\infty}^\infty\;d\rho
(2\pi)^{-1}|\varphi_j(\rho,\zeta,\epsilon)|^2$ and that 
$\varphi_j(\rho,\zeta,\epsilon)=\int\limits_0^\infty\;d\ln(r)e^{is\ln(r)}
g_j(r,\zeta,\epsilon)$. Hence, upon integration over the compact
$S^2$-coordinates we recover \eqref{int1} and \eqref{int2}. The overall
construction relies only on Mellin inversion formula and the Plancherel
theorem; hence the map from $F_j$ onto $\varphi_j$ exists whenever
the latter is square-integrable; this concludes the demonstration.
\end{proof}

To conclude the analysis on the functions over a light cone, let us
recall the following result still from \cite{Strichartz2}:\\

\lemma{\em Whenever $\rho\neq 0$ then 
\begin{gather}\label{A0}
A_0(\rho)\varphi(\zeta^\prime,\epsilon^\prime)=\frac{\rho}{\pi}\sume\int
\limits_{\bS^2}|\vec{\zeta}\cdot\vec{\zeta}^\prime-\epsilon\epsilon^\prime|^{-1-i
\rho}\varphi(\zeta,\epsilon)d\zeta,\\
A_1(\rho)\varphi(\zeta^\prime,\epsilon^\prime)=\frac{\rho}{\pi}\sume\int
\limits_{\bS^2}|\vec{\zeta}\cdot\vec{\zeta}^\prime-\epsilon\epsilon^\prime|^{-1-i
\rho}sgn\left(\vec{\zeta}\cdot\vec{\zeta}^\prime-\epsilon\epsilon^\prime\right)
\varphi(\zeta,\epsilon)d\zeta,\label{A1}
\end{gather}
are unitary operators respectively on odd and on even functions in
$L^2(\bS^2\times\pm 1) $. In \eqref{A0} and \eqref{A1} $``\cdot"$
stands for the standard Euclidean scalar product on $\bR^3$, whereas a 
function $f(\zeta,\epsilon)\in L^2(\bS^2\times\pm 1)$ iff 
$$\sume\int\limits_{\bS^2}|f(\zeta,\epsilon)|^2<\infty.$$}

We can now put together the previous lemma and proposition \ref{prop1}
in order to represent any function $F\in L^2(\mC)$ as

\begin{gather}F(p_\mu)=\frac{1}{2\pi^3}\sumep\int\limits_{-\infty}^\infty
d\rho\;\rho^2\int\limits_{S^2}d\zeta^\prime|\vec{p}\cdot\vec{\zeta}^
\prime-p_0\epsilon^\prime|^{-1+i\rho}\left[\psi_0(\rho,\zeta^\prime,
\epsilon^\prime)\right.+\notag\\
\label{decomp}
+\left.\psi_1(\rho,\zeta^\prime,\epsilon^\prime)sgn(\vec{p}\cdot\vec{
\zeta}^\prime-p_0\epsilon^\prime)\right], 
\end{gather}
being $\psi_k(\rho,\zeta^\prime,\epsilon^\prime)\doteq\frac{\pi}{\rho}
A_k(\rho)\varphi_k(\rho,\zeta^\prime,\epsilon^\prime)$ with $k=0,1$
and $\varphi_k$ chosen according to \eqref{int2}.

\vskip .5cm

\noindent Let us now move back to the square integrable functions over 
$\bH_m$ and, to fix notations, let us call 
$\widetilde\square=-\frac{\partial^2}{\partial p_0^2}+\sum\limits_{i=1}^3
\frac{\partial^2}{\partial p^2_i}$. Switching from the coordinates
$(p_0,p_i)$ to $(m,r,\zeta)$ as introduced at the beginning of this 
section, the D'Alambert operator becomes $\widetilde\square=-\frac{
\partial^2}{\partial m^2}-\frac{3}{m}\frac{\partial}{\partial m}+\frac{
\square_{\bH}}{m^2}$, where $\square_{\bH}$ is the Laplacian on the 
unit hyperboloid. 

It is standard result that $\square_{\bH}$ is a selfadjoint operator on
$\left\{f\in L^2(\bH_m),\;|\;\square_{\bH}f\in L^2(\bH_m)\right\}$ with
a continuous negative spectrum; furthermore it commutes with the 
quasi-regular $O(3,1)$ representation i.e. 
$[U(\Lambda),\square_{\bH}]=[U(\Lambda),\widetilde\square]=0$ for any
$\Lambda\in O(3,1)$.

The strategy is to consider the mass
hyperboloid as a non characteristic initial surface for the wave
equation $\widetilde\square u(m,r,\zeta,\epsilon)=0$ to be solved in the
region $m\geq 0$. In particular the following lemma holds:\\

\lemma{\label{massless}\em Calling $B=-\square_{\bH}-1$, then for any $f,g\in
L^2(\bH_m)$ the function 
\beq\label{equm}
u(m,r,\zeta,\epsilon)=m^{-1+i\sqrt{B}}f(r,\zeta,\epsilon)+m^{-1-i
\sqrt{B}}g(r,\zeta,\epsilon)
\eeq
satisfies $\widetilde\square u=0$ for $m>0$ with Cauchy data
$$u(1,r,\zeta,\epsilon)=f(r,\zeta,\epsilon)+g(r,\zeta,\epsilon),\quad
iB^{-\frac{1}{2}}\frac{\partial (mu)}{\partial m}(1,r,\zeta,\epsilon)=g
(r,\zeta,\epsilon)-f(r,\zeta,\epsilon).$$
Furthermore for all $m>0$ it holds
$$2\left(||f||^2_2+||g||^2_2\right)=\int\limits_{\bH_m}
d\mu(\bH_m)m^2\left(|u(m,r,\zeta,\epsilon)|^2+\left|B^{-\frac{1}{2}}
\frac{\partial(mu)}{\partial m}(r,\zeta,\epsilon)\right|^2\right),$$
where $||,||_2$ is the norm \eqref{norm} with $\alpha=2$.}\\ 

\begin{proof}
If we show that $u(m,r,\zeta,\epsilon)$ is a solution of D'Alambert wave 
equation then the statement on Cauchy data holds per direct substitution 
and the identity between norms stands per unitarity of the operator 
$m^{i\sqrt{B}}$ on $L^2(\bH_m)$. 

Hence let us consider any but fixed $v\in C^\infty_0(\bR^4)$ whose 
support does not include the origin. Dropping the $\epsilon$ dependence 
which is irrelevant to the proof, integration per parts grants:
$$\int\limits_{\bR^4}d^4p\; v(p^\mu)\widetilde\square u(p^\mu)=
\int\limits_{\bR^4}d^4p\; u(p^\mu)\widetilde\square v(p^\mu).$$
In terms of coordinates $(m,r,\zeta)$ this last identity reads
\begin{gather*}\int\limits_{\bR^4}dm\;d\mu(\bH_m)\;m^3u(m,r,\zeta)\left(
-\frac{\partial^2}{\partial m^2}-\frac{3}{m}\frac{\partial}{\partial m}+
\frac{\square_{\bH}}{m^2}\right)v(m,r,\zeta)=\\
\int\limits_{\bR^4}dm\;d\mu(\bH_m)\;m v(m,r,\zeta)\left(
\square_{\bH}-m^2\frac{\partial^2}{\partial
m^2}-3m\frac{\partial}{\partial m}\right)u(m,r,\zeta),
\end{gather*}
which, inserting the expression for $u(m,r,\zeta)$ in the hypothesis,
becomes
$$\int\limits_{\bR^4}dm\;d\mu(\bH_m)\;m
v(m,r,\zeta)\left(\square_{\bH}+1+B\right)u(m,r,\zeta)=0,$$
being $B\doteq -\square_{\bH}-1$.
\end{proof}

The choice of the initial surface as the unitary hyperboloid is pure
convenience and no generality is lost in this process since it is
possible to pick any $\bH_m$ and none of the forthcoming results would
be modified. The independence from $m$ in the norm identity in the last lemma 
and the equality $\lim_{m\to 0}m^2d\mu(\bH_m)=d\mu(\mC)$ suggests that 
we are now in position to construct a unitary intertwining operator
$\widetilde T:L^2(\mC)\to L^2(\bH_m)\oplus L^2(\bH_m)$. 
As a matter of fact all the needed ingredients can be found in the 
previous lemma and in formula \eqref{decomp}:\\

\proposizione\label{ultima}{\em
Given any function $F\in L^2(\mC)$, let us decompose it as $F=F_+ + F_-$
where $+$ represents the contribution of the integral in the
$\rho$-variable between $0$ and infinity in \eqref{decomp} whereas the
pedex $-$ refers to that between minus infinity and $0$. Then, 
if $f\doteq F_+|_{\bH_m}$ and $g\doteq F_-|_{\bH_m}$, the function $u$
constructed as in lemma \ref{massless} coincides with $F$. Furthermore
the map from $F|_\mC\longrightarrow L^2(\bH_m)\oplus L^2(\bH_m)$ is an
intertwiner between the $O(3,1)$ representations.}\\

\noindent The demonstration is left to \cite{Strichartz2}.\\

\remark{A consequence of the above proposition is that any 
$f\in L^2(\bH_m)$ can be decomposed as 
\beq\label{intert}
f(p^\mu)=\sume\int\limits_{0}^\infty
\frac{d\rho}{2\pi^3}\;\rho^2\int\limits_{\bS^2}d\zeta^\prime|\vec{p}
\cdot\vec{\zeta}^\prime-\epsilon
E|^{-1+i\rho}\left[\psi_0(\rho,\zeta^\prime,\epsilon)+sgn\left(\vec{p}
\cdot\vec{\zeta}^\prime-\epsilon p_0\right)\psi_1(\rho,\zeta^\prime,\epsilon)
\right],
\eeq 
where 
$$\psi_0(\rho,\zeta^\prime,\epsilon)=\int\limits_{\bR^4}d^4p\;\delta(\eta
^{\mu\nu}p_\mu p_\nu-m^2)f(p^\mu)|\vec{p}\cdot\vec{\zeta}^\prime-
\epsilon p_0|^{-1-i\rho},$$
and 
$$\psi_1(\rho,\zeta^\prime,\epsilon)=\int\limits_{\bR^4}d^4p\;\delta(\eta
^{\mu\nu}p_\mu p_\nu-m^2)f(p^\mu)|\vec{p}\cdot\vec{\zeta}^\prime-
\epsilon E|^{-1-i\rho}sgn\left(|\vec{p}\cdot\vec{\zeta}^\prime-\epsilon 
p_0|\right).$$

\noindent Let us pinpoint
\begin{enumerate}
\item although \eqref{intert} is written in terms of the global 
coordinates, we can switch to intrinsic coordinates $(r,\zeta,\epsilon)$
over $\bH_m$ simply substituting $\vec{p}$ with $\vec{\zeta}$ and $p_0$ 
with $\epsilon$.
In other words we have decomposed a generic function $f\in L^2(\bH_m)$ into a
direct integral in terms of irreducible representations of $O(3,1)$.

\item proposition \ref{ultima} provides a way to explicitly construct
the inverse intertwiner $T=\widetilde T^{-1}: L^2(\bH_m)\oplus L^2(\bH_m)
\to L^2(\mC)$. As a matter of fact starting from any two functions 
$f,g\in L^2(\bH_m)$, one can generate a solution of D'Alambert wave
equation out of \eqref{equm} whose restriction to the light cone is a
function $F\in L^2(\mC)$; in a few words $T(f,g)=F$.

From our perspective this a slightly inconvenient situation since we
start with a solution of \eqref{KG} and hence with a single function
$f\in L^2(\bH_m)$. Unfortunately the Cauchy problem, upon which
\eqref{equm} is based, requires two initial  condition. Hence we adopt
the choice to imbed $L^2(\bH_m)$ into the diagonal component of 
$L^2(\bH_m)\oplus L^2(\bH_m)$, namely we fix the map $i:L^2(\bH_m)\to
L^2(\bH_m)\oplus L^2(\bH_m)$ such that $i(f)=(f,f)$. Clearly this choice
is not unique and the resulting function on the light cone we will
construct depends also upon the choice of $i$.

To summarise we have set the map $T\circ i:L^2(\bH_m)\to L^2(\mC)$
such that $T\left(i(f)\right)=F$.
\end{enumerate}}

In order to complete our task, a last question must be answered namely
if, to any element of $L^2(\mC)$, it corresponds a function in Minkowski
spacetime  which solves the wave equation. A positive answer has been
already given in \cite{Strichartz3} and, thus, we end up with:\\

\proposizione\label{Stri}{\em
If $F(r,\zeta,\epsilon)\in L^2(\mC)$, then it is the restriction on the
light cone of the Fourier transform of a function 
$\psi\in L^4(M^4,d^4x)$ which solves the wave
equation $\square_\eta\psi(x^\mu)=0$ with Cauchy data 
$$\psi(0,x^i)=f_1(x^i),\quad\frac{\partial\psi}{\partial
t}\left(0,x^i\right)=f_2(x^i),$$
with $K^{\frac{1}{2}}f_1(x^i)$ and $K^{-\frac{1}{2}}f_2(x^i)\in L^2(
\bR^3, d^3 x)$ (j=1,2) where $K=\sqrt{-\triangle}$ and 
$\triangle=\sum\limits_{i=1}^3\frac{\partial^2} {\partial x_i^2}$.

Furthermore it exists a suitable constant $C$ such that 

\beq\label{normestim}
||\psi(x^\mu)||_{L^4(M^4)}\leq C\left(||K^{\frac{1}{2}}f_1(x^i)||_{L^2(
\bR^3)}+||K^{-\frac{1}{2}}f_2(x^i)||_{L^2(\bR^3)}\right).
\eeq}

\begin{proof}
The first part of the proposition is proved in lemma 1 of Strichartz
seminal paper \cite{Strichartz3}. Hence we know that $\phi(x^\mu)$ is a
solution for the wave equation lying in $L^4(M^4,d^4x)$ and we need only 
to focus on Cauchy data. In a standard Minkowski frame with 
coordinates $x^\mu=(t,\vec{x})\in\bR^4$ we can decompose the solution 
for the wave equation constructed out of $F$ as
$$\psi(t,\vec{x})=\int\limits_{\bR^3}\frac{d^3p}{\sqrt{16\pi^3|}\vec{p}|}
\left[e^{i\left(\vec{p}\cdot\vec{x}-t|\vec{p}|\right)}F_+(p)+e^{i\left(
\vec{p}\cdot\vec{x}+t|\vec{p}|\right)}F_-(p)\right],$$
where $F_+$ and $F_-$ are respectively the restriction of $F$ to the
upper and lower light cone. Taking into account the identity
$$\psi(t,\vec{x})=-i\int\limits_{\bR^3}\frac{d^3p}{\sqrt{16\pi^3}}
K^{-\frac{1}{2}}\left[e^{i\left(\vec{p}\cdot\vec{x}-t|\vec{p}|\right)}
\frac{F_+(p)}{\sqrt{|\vec{p}|}}+e^{i\left(\vec{p}\cdot\vec{x}+t|\vec{p}|
\right)}\frac{F_-(p)}{\sqrt{|\vec{p}|}}\right],$$
and evaluating this expression for $t=0$ we discover that
$K^{\frac{1}{2}}\psi(0,\vec{x})$ is up to a multiplicative constant
complex number the sum of the Fourier transform of
$\frac{F_\pm(p)}{\sqrt{2|\vec{p}|}}$; hence, being $F\in L^2(\mC)$, per
Plancherel theorem $K^{\frac{1}{2}}\psi(0,\vec{x})\in L^2(\bR^3,d^3x)$.

Deriving now once in the time variable and exploiting the same kind of
identity, we end up with
$$\frac{\partial\psi}{\partial t}(t,\vec{x})=i\int\limits_{\bR^3}\frac{
d^3p}{\sqrt{16\pi^3}}K^{\frac{1}{2}}\left[e^{i\left(\vec{p}\cdot\vec{x}
-t|\vec{p}|\right)}\frac{F_+(p)}{\sqrt{|\vec{p}|}}-e^{i\left(\vec{p}
\cdot\vec{x}+t|\vec{p}|\right)}\frac{F_-(p)}{\sqrt{|\vec{p}|}}\right].$$
Hence, evaluating at $t=0$ this expression and still exploiting the
Plancherel theorem as in the previous case, we end up with $K^{\frac{1}
{2}}\frac{\partial\psi}{\partial t}(0,\vec{x})\in L^2(\bR^3,d^3x)$.

To conclude the demonstration it suffices to notice that the field 
$\psi(x^\mu)$ and the functions $f_1(x^i),f_2(x^i)$ satisfy the 
hypotheses of corollary 2 in \cite{Strichartz3} where the norm estimates 
\eqref{normestim} for the homogeneous D'Alambert wave equation have been 
proved.
\end{proof}

\remark{On an operative ground the solution of the D'Alambert wave
equation can be constructed starting from any but fixed 
$f\in L^2(\bH_m)$, map in $F=T\left(i(f)\right)\in L^2(\mC)$, decompose
it as in \eqref{decomp} and eventually perform an inverse Fourier 
transform {\it i.e} 
\begin{gather}\notag
\psi(x^\mu)=\int\limits_{\bM^4}\frac{d^4x}{4\pi^2}\;
e^{i\eta^{\mu\nu}x_\mu p_\nu}\sume\int\limits_{-\infty}^\infty
\frac{d\rho}{2\pi^3}\;\rho^2\int\limits_{\bS^2}d\zeta^\prime|\vec{p}
\cdot\vec{\zeta}^\prime-\epsilon
p_0|^{-1+i\rho}\left[\psi_0(\rho,\zeta^\prime,\epsilon)+\right.\\
+\left.sgn\left(\vec{p}
\cdot\vec{\zeta}^\prime-\epsilon p_0\right)\psi_1(\rho,\zeta^\prime,\epsilon)
\right].\label{massless2}
\end{gather}}

\sssa{From bulk to null infinity}

The results from the previous section can be applied to introduce a 
``projection'' of finite-norm solutions $\phi$ for the massive 
Klein-Gordon equation to null infinity. In particular let us summarise
that all the informations of $\phi$ can be encoded in the following 
triplet of data:
\begin{enumerate}
\item the function $\psi(x^\mu)$ constructed as in \eqref{massless2}
which solves the massless Klein-Gordon equation of motion along the
lines of proposition \ref{Stri},
\item the quasi-regular representation $U^\prime(\Lambda)$,
\item the intertwiner $\widetilde T:L^2(\mC)\to L^2(\bH_m)\oplus
L^2(\bH_m)$. 
\end{enumerate}

\noindent Thus the overall problem reduces to find a projection for
$\psi(x^\mu)$ to null infinity. 

As a first step let us remember that Minkowski spacetime can be 
compactified in the Einstein static universe \cite{Wald}. More in detail, 
let us consider the coordinates $(u,v,\theta,\varphi)$ being $(\theta,
\varphi)$ the standard coordinates on $\bS^2$, $u=t+r$ and $v=t-r$ with 
$r$ as radial coordinate and let us choose as conformal factor 
\beq\label{conf}
\Omega^2=4\left[(1+u^2)(1+v^2)\right]^{-1}.
\eeq 
Hence the flat metric is rescaled to 
$$ds^{\prime 2}=\widehat g^{\mu\nu}dx_\mu
dx_\nu=\frac{4}{(1+u^2)(1+v^2)}\left[-dudv+\frac{(u-v)^2}{4}d\bS^2(\theta
,\varphi)\right],$$
with $d\bS^2(\theta,\varphi)\doteq d\theta^2+\sin^2\theta d\varphi^2$.
If we perform the change of variables 
\beq\label{change}
T=\tan^{-1}u+\tan^{-1}v,\quad R=\tan^{-1}u-\tan^{-1}v,
\eeq
then we can realize the original Minkowski spacetime as the locus 
$\left(-\pi,\pi\right)\times\left(-\pi,\pi\right)\times S^2\subset\bR
\times S^3$ with respect to the metric
\beq\label{Einst}
ds^{\prime 2}=\widehat g^{\mu\nu}dx_\mu dx_\nu=-dT^2+dR^2+\sin^2R\;
d\bS^2(\theta,\varphi),
\eeq
{\it i.e.} that of Einstein static universe $\widehat M$. 
Let us notice that, the closure of the image of Minkowski spacetime in
$(\bR\times \bS^3,\widehat{g}_{\mu\nu})$ is compact and that $\Im^+$ is 
nothing but the locus $T+R=\pi$. 

More importantly this new background in still globally hyperbolic
and, if we introduce $\widetilde\psi\doteq\Omega^{-1}\psi$, then it is a
solution of the Klein-Gordon equation $\square_{\widehat
g}\phi-\frac{\widehat{R}}{6}\phi=0$ where $\square_{\widehat g}\doteq 
\widehat{g}^{\mu\nu}\widehat{\nabla}_\mu\widehat{\nabla}_\nu$ is the 
wave operator with respect to the metric $\widehat g_{\mu\nu}$ and
$\widehat{R}=1$ is the scalar curvature of Einstein static
universe. Furthermore, since the original Cauchy surface $\bR^3$ at 
$t=0$ is mapped into $T=0$ in $\widehat M$ we can recast the Cauchy 
problem in proposition \ref{Stri} as 
\beq\label{new}
\left\{\begin{array}{l}
\square_{\widehat g}\widetilde{\psi}(X^\mu)=\frac{\widetilde\psi(X^\mu)}{6}\\
\widetilde{\psi}(0,X^i)=f_1(X^i)\\
\frac{\partial\widetilde\psi}{\partial
T}\left(0,X^i\right)=f_2(X^i)
\end{array}
\right.,
\eeq
where $X^\mu\doteq(T,X^i)=(T,R,\theta,\varphi)$ and where $\widetilde{K}
^{\frac{1}{2}}f_1(X^i)\in L^2(\bS^3)$ and $\widetilde{K}^{-\frac{1}{2}}f
_2(X^i)\in L^2(\bS^3)$ being $\widetilde{K}$ the square-root of the 
Laplace-Beltrami operator out of the spatial component of the metric 
\eqref{Einst}. Here square integrability is meant with respect to the 
measure $d\mu=\sin^2R\sin\theta dRd\theta d\varphi$.

Hence $\widetilde\psi(X^\mu)$ satisfies the Klein-Gordon equation 
with $m^2=\frac{1}{6}$, it coincides with $\Omega^{-1}\psi$ in the image 
of Minkowski spacetime in $\widehat M$ and furthermore it lies in
$L^4(M^4,\sqrt{|\widehat{g}|}d^4X)$ since 
$$||\psi(x^\mu)||^4_{L^4}=\int\limits_{\bR^4}|\psi(x^\mu)|^4d^4x=\int
\limits_{\bR^4}|\widetilde\psi(x^\mu)|^4\Omega^4d^4x=\int
\limits_{-\pi}^\pi\int\limits_{-\pi}^\pi\int\limits_{\bS^2} d^4X\sqrt{|
\widehat{g}|}|\widetilde\psi(X^\mu)|^4,$$
where in the last equality we exploited the coordinate change
\eqref{change}. 

Unfortunately, since our aim is to project $\widetilde\psi$ on null
infinity, the best available tools to define a function on $\Im^+$ are
trace theorems for Sobolev spaces. In order to exploit them the set of 
solutions for the wave equation we are taking into
account is too big and thus we need to consider only more regular
solutions for the wave equation.

To understand which is the less restrictive constraint we have to
impose, let us gather all the needed ingredients. As a first step we 
point out that, being Minkowski spacetime an open set of finite volume 
(either with respect to Lesbegue measure or with respect to
$\sqrt{|\widehat{g}|}dTdRd\bS^2(\theta,\varphi)$) in Einstein static
universe, then H\"older inequality grants us that $L^p(M)\subset L^q(M)$ 
for all $1\leq q< p\leq\infty$. This property can be recast at a level 
of first order Sobolev spaces in $L^p(M)$ {\it i.e.}
$W^{1,p}(M)\subset W^{1,q}(M)$ for $1\leq q< p\leq\infty$.

As a second step we aim to exploit proposition 4.3 in \cite{Showalter} 
according to which, if $\Omega$ is a bounded domain in $\bR^N$ with a 
three dimensional $C^1$-boundary $\partial\Omega$, then it exists a 
linear trace operator $\gamma:W^{1,p}(\Omega)\to L^p(\partial\Omega)$ 
which is  continuous and uniquely determined by the boundary value of 
the functions $u\in C^1(\overline{\Omega})$. Furthermore the kernel of
$\gamma$ is $W^{1,p}_0(\Omega)$ {\it i.e.} the closure of $C^\infty_0(
\Omega)$ in $W^{1,p}(\Omega)$. 

Our scenario meets all the geometric requirements in the
above hypothesis since Minkowski background is a bounded open set in
Einstein static universe $\bR\times\bS^3$ which, in its turn, can be 
identified as an open set of $\bR^5$. Furthermore 
the boundary of $M$ consists of two smooth null hypersurfaces - future 
and past null infinity - and thus, taking into account that
$\widetilde\psi$ lies in $L^4(M, d^4X)$ and hence in $L^p(M, d^4X)$ for
all $1\leq p\leq 4$,  we can apply such proposition only to those 
$\widetilde\psi\in W^{1,p}(M)$ still with $1\leq p\leq 4$. 

With this further condition set and with the inclusion relations between 
the Sobolev spaces as discussed before, we are entitled to introduce the
the map $\gamma|_{\Im^+}:W^{1,p}(M^4)\to L^q(\Im^+)$ where $q$ can be
fixed to any value lower or equal to $p$. Here $\Im^+$ is the locus
$(-\pi,\pi)\times\bS^2$ and the measure on $\Im^+$ is the Lesbegue one.
Hence being $\Im^+$ in this reference frame an open set of
$\bR\times\bS^2$, each function on $L^q(\Im^+)$ can be also read as an
element in $L^q(\bR\times S^2)$. This property will be exploited in the
next section.

Taking into account that, both from a physical point of view and for the
analysis in the next section, it is better to work with Hilbert spaces
on the boundary we can summarise the previous discussion as:\\

\proposizione\label{proj}{\em Assume that Minkowski spacetime $M$ is 
conformally embedded as an open set of Einstein static universe 
$\left(\widehat M, \widehat g\right)$ with $\widehat g$ as in 
\eqref{Einst}. Then, for any solution of the wave equation $\psi\in L^4(
\bR^4, d^4x)$ the function $\widetilde\psi\doteq\Omega^{-1}\psi\in L^4(\widehat 
M,\sqrt{|\widehat g|}d^4x)$ - with $\Omega$ chosen as in \eqref{conf} - 
solves \eqref{new}. Furthermore, whenever $\widetilde\psi\in W^{1,p}(M)$ with 
$p\leq 4$, it exists a continuous projection operator $\gamma|_{\Im^
+}:W^{1,p}(M)\to L^q(\Im^+)$ where we fix $q=2$ if $2\leq p\leq 4$ 
whereas $q=1$ if $p=1$. The image $\Psi$ under $\gamma|_{\Im^+}$ of 
$\widetilde\psi$ will be referred to as its restriction on future null 
infinity.}\\

\remark\label{smooth}{This last proposition partly overlaps the scenarios 
envisaged in \cite{Dappiaggi, Mason} where only solutions $\psi$ to the 
D'Alambert wave equation with compactly supported initial data where 
taken into account. As partly discussed in the introduction, in this case, 
$\psi\in C^\infty(\bR^4)$ and accordingly also $\widetilde\psi\in 
C^\infty(M)$ adopting the nomenclature of the previous analysis. 
Furthermore, the uniqueness of the solution for the Cauchy problem of the 
Klein-Gordon equation in the Einstein static universe allows to construct 
a unique function in $\widehat M$ coinciding  with $\widetilde\psi$ if 
restricted to $M$. Hence, in this case, restriction to $\Im^+$ simply 
means the evaluation of the solution on future null infinity.}\\

\remark{We point out that the additional regularity condition ({\it
i.e.} $\widetilde\psi\in W^{1,p}(M)$ on the solutions for the D'Alambert 
wave equation haw been set in the Einstein static universe because a 
direct inspection of the previous construction shows
that, although, whenever $f\in L^p(\bR^4,d^4x)$, $\Omega^{-1}f\in
L^p(\bR^4,\sqrt{|\widehat{g}|}d^4x)$ for $p\leq 4$, this does not held
true for first order Sobolev spaces. In other words $f\in
W^{1,p}(\bR^4)$, then, exploiting Liebinitz rule, one can realize that, 
due to the contribution of the derivatives of the conformal factor 
\eqref{conf}, $\Omega^{-1}f\in L^p(\bR^4,\sqrt{|\widehat{g}|}d^4x)$ but 
not necessary in $W^{1,p}(\bR^4, \sqrt{|\widehat{g}|}d^4x)$.}\\

Hence we have achieved our goal since all the information from the 
original massive field $\phi$ satisfying \eqref{KG} has been projected 
onto null infinity in the triplet $(\Psi, U^\prime, T)$ where $U^\prime$
is the quasi-regular $O(3,1)$ representation acting on the massless 
field and $T$ is the intertwiner constructed in the previous section. 
Two natural questions arise at this stage:
\begin{itemize}
\item What about Poincar\'e covariance?
\item What is the field theoretical meaning that $(\Psi, U^\prime, T)$ 
contains the information of the massive scalar field?
\end{itemize}

Let us answer to the first and simpler question. Up to now we have
considered only the quasi-regular $O(3,1)$ action on the set $L^2(\mC)$
or $L^2(\bH_m)$. If we want to deal instead with Poincar\'e covariant
scalar field theories, a function $\phi$ satisfying either \eqref{KG} or 
D'Alambert wave equation would transform in a momentum frame as
$$\widetilde{U}(\Lambda,a^\mu)\hat\phi(p_\mu)=e^{ia^\mu
p_\mu}\hat\phi(\Lambda^{-1}p_\mu),$$
where the hat symbol stands for the Fourier transform. This identity
supplemented with the constraints $\eta^{\rho\sigma}p_\rho p_\sigma\phi(
p_\mu)=m^2\phi(p_\mu)$ with $m^2\geq 0$ and $sgn(p_0)>0$ is a unitary 
irreducible representation for the full Poincar\'e group \cite{Group}. 

In order to relate the two above points of view, beside the trivial 
restriction from $O(3,1)$ to $SO(3,1)$, we need only to invoke the
induction-reduction theorem ({\it c.f.} chapter 18 in \cite{Group})
according to which the quasi-regular representation $U(\Lambda)$ on
$L^2(\bH_m)$ is\\ 
a) the $SO(3,1)$ representation induced from the identity representation 
of $SO(3)$,\\
b) the restriction of the scalar Poincar\'e representation to the
Lorentz group. At the same time, if we start from $U(\Lambda)$, it 
induces the unitary and irreducible scalar representation of the full 
Poincar\'e group.

A similar reasoning and conclusion holds if we consider $L^2(\mC)$ with 
the associated quasi-regular representation $U^\prime(\Lambda)$.\\
  
\ssa{Data reconstruction on null infinity}
In this last subsection we face the last and most important question
namely in which sense the information from the bulk massive field
projected on null infinity out of $(\widetilde\psi, U^\prime, T)$ can be
interpreted from a classical field theory perspective. 
To this end we shall exploit some recent analysis according to which it
is possible to explicitly construct a diffeomorphism invariant field 
theory on future null infinity. Afterwards our aim will be to show how 
the above triplet can be interpreted in terms of such a boundary free 
field theory. 

Bearing in mind the notations and the nomenclatures of subsection 
\ref{AFS}, we review some feature of  the construction of a Poincar\'e 
invariant field theory on $\Im^+$ - thought as a null differentiable 
manifold\footnote{More
appropriately one should claim that we are constructing a QFT on 
the equivalence class of triplets $(\Im^+,n^a, h_{ab})$ associated to
the bulk Minkowski spacetime.} - for smooth
scalar fields invariant under the $\mathcal{R}$ subgroup of the BMS as
discussed in \cite{Arcioni, Dappiaggi, Dappiaggi2}. 
Such problem has been discussed for the full $SO(3,1)\ltimes
C^\infty(\bS^2)$; hence here we will adapt that analysis to the specific
scenario of bulk Minkowski background.

To this end we shall follow two possible roads: the first starts from a
massless bulk scalar field and it imposes BMS invariance on the smooth
projection of such a field on null infinity whereas the second ignores
the bulk and it constructs a scalar free field theory on $\Im^+$ by means 
of the Mackey-Wigner programme {\it i.e.} we only exploit the knowledge 
of the symmetry group.

We stress that the full construction has been developed for a generic 
asymptotically flat spacetime due to the universality of the boundary 
structure. Hence, although both the above mentioned approaches have been 
fully accounted for in \cite{Dappiaggi, Dappiaggi2}, here we will only 
review the details adapted to the case of Minkowski bulk spacetime and, 
thus, Poincar\'e symmetry group on null infinity leaving an interested 
reader to the above cited manuscripts for a careful analysis.

Let us thus start from the first part of this programme; in order to
construct a meaningful scalar field theory on $\Im^+$ starting from the 
bulk, we can focus only smooth real solutions $\psi$ for the D'Alambert 
wave equation. As per remark \ref{smooth} such a bulk field projects to 
$\Psi\in C^\infty(\Im^+)$.
Then, if we wish to define a suitable representation of $\mathcal{R}$
acting on each $\Psi$, the following proposition holds \cite{Dappiaggi}:
\\

\proposizione{\em Let us take Minkowski spacetime $(M^4,\eta_{\mu\nu})$ 
and an associated compactified spacetime $(\widehat M,\widehat g_{\mu\nu}
)$ (not necessarily Einstein static universe) and let us fix an arbitrary 
gauge factor $\omega$. Then, for any but fixed $\lambda\in\bR$ and for 
any but fixed $g\in\mathcal{R}\subset BMS$, a representation is 
$A^{(\lambda)}(g):C^\infty(\Im^+)\to C^\infty(\Im^+)$ such that the map 
$t\mapsto A^{(\lambda)}(g_t)\Psi=\lim\limits_{\Im^+}\left(\omega\Omega
\right)^\lambda g^*_t(\widetilde\psi)$ is smooth for every fixed bulk 
scalar field $\psi$ with smooth projection $\Psi$ on $\Im^+$ and for 
every but fixed one-parameter subgroup of the bulk Poincar\'e group. In 
the Bondi  frame $(u,z,\bar{z})$ it reads
$$\left(A^{(\lambda)}(g)\Psi\right)(u^\prime,z^\prime,\bar{z}^\prime)=
K_{\Lambda}^{-\lambda}(z,\bar{z})\Psi(u,z,\bar{z}),\quad\forall
g=\left(\Lambda,\alpha(z,\bar{z})\right)\in\mathcal{R}$$
where the primed coordinates and $K_{\Lambda}(z,\bar{z})$ are defined as
in \eqref{BMS1} and \eqref{BMS2}.}\\

Since our aim is to deal with unitary and irreducible representations we
have to go one step further {\it i.e.}\\

\proposizione\label{omni}{\em Let us consider the set $\mS(\Im^+)\subset
C^\infty(\Im^+)$ of real functions $\Psi$ such that $\Psi$ itself and
all its derivatives decay faster than any power of $|u|$ when
$|u|\to\infty$ and uniformly in $(z,\bar{z})$. Then $\mS(\Im^+)$ can be
endowed with the strongly non degenerate symplectic form
$$\sigma(\Psi_1,\Psi_2)=\int\limits_{\bR\times\bS^2}\left(\Psi_2\frac{
\partial\Psi_1}{\partial u}-\Psi_1\frac{\partial\Psi_2}{\partial
u}\right)dud\bS^2(z,\bar{z}),$$
and $(\mS(\Im^+),\sigma)$ is invariant only under $A^{(1)}(g)$.
Furthermore if we introduce the positive frequency part $\widehat\Psi_+$ 
of $\Psi\in\mS(\Im^+)$ as 
\beq\label{Four}
\widehat\Psi_+(E,z,\bar{z})=\int\limits_\bR \frac{du}{\sqrt{2\pi}}\;e^{
iEu}\Psi(u,z,\bar{z}),\quad E\in[0,\infty)
\eeq
we can write $\widehat\Psi=\widehat\Psi_++\overline{\widehat\Psi_+}$. If 
we denote with $\mS(\Im^+)_{\bC}$ the complex linear combinations of 
these functions $\widehat\Psi(E,z,\bar{z})$ then 
\begin{enumerate}
\item $\mS(\Im^+)_{\bC}$ can be closed to Hilbert space $\mch$ with
respect to the Hermitian inner product 
$$\langle\widehat\Psi_1,\widehat\Psi_2\rangle=-i\sigma(\overline{\widehat
\Psi_1},\widehat\Psi_2).$$
Furthermore $(\mch,\langle,\rangle)$ is unitary isomorphic to
$L^2(\bR\times\bS^2,EdEd\bS^2(z,\bar{z}))$
\item the representation $A^{(1)}(g)$ of $\mathcal{R}$ on $\mch$ acts
as
\beq\label{BMS}
\left(A^{(1)}(g)\widehat\Psi\right)(E,z,\bar{z})=e^{iEK_\Lambda(\Lambda^
{-1}z,\Lambda^{-1}\bar{z})\alpha(z,\bar{z})}\widehat\Psi(EK_\Lambda(
\Lambda^{-1}z,\Lambda^{-1}\bar{z}),\Lambda^{-1}z,\Lambda^{-1}\bar{z}),
\eeq
for any $g=(\Lambda,\alpha(z,\bar{z}))\in\mathcal{R}$ and $A^{(1)}(g)$ is
unitary on $\mch$.
\end{enumerate}
}

\noindent The proof of this theorem is a recollection with minor
modifications of the demonstration of proposition 2.9, 2.12 and 2.14 in 
\cite{Dappiaggi}. Hence we refer to such paper for an interested
reader.\\

\noindent We now state a useful lemma out of this last proposition:\\

\lemma\label{un1}{\em The projection on $\Im^+$ of each function 
$\widetilde\psi$ constructed as in proposition \ref{proj} can be unitary 
mapped into an element of $(\mch,\langle,\rangle)$.}

\begin{proof}
In proposition \ref{proj} we projected a function with support on the 
image of Minkowski spacetime in Einstein static universe to a function $\Psi\in
L^2(\Im^+)$ being $\Im^+$, in that specific background, $(-\pi,\pi)\times
\bS^2$. Since $\bS^2$ is compact and $(-\pi,\pi)$ is an open bounded set 
of $\bR$, $\Psi$ can also be read as an element of $L^2(\bR\times\bS^2)$. 
We stress that, switching from the Lesbegue measure in $L^2(\Im^+)$ to 
the natural $SO(3)$-invariant measure on $\bS^2$ for $L^2(\bR\times
\bS^2)$ is harmless. 

According to Plancherel theorem and to \eqref{Four} the Fourier
transform $\widehat\Psi\in L^2(\bR\times S^2,EdEd\bS^2)$ and, hence,
according to proposition \ref{omni}, it can be unitary mapped in $(\mch,
\langle,\rangle)$.  
\end{proof}

This concludes the first part of our programme though a complete
analysis would require the proof that $A^{(1)}$ is irreducible or how it
decomposes in irreducible components. The answer to this question will
be a byproduct of the Wigner-Mackey analysis that we discuss now.\\ 
Such approach calls for the construction of a classical free field theory on
a generic manifold only by means of the symmetry group,
$\mathcal{R}\subset BMS$ in our case. Although $\mathcal{R}$ is
homomorphic to the Poincar\`e group we cannot simply refer to the
standard construction for a covariant field theory in Minkowski
background as discussed to quote just one example in chapter 21 of
\cite{Group}. On the opposite we need to consider $\mathcal{R}$ as a
subgroup of the BMS and hence we shall adapt the analysis in 
\cite{Dappiaggi} to this simpler scenario. 

Referring to this last cited paper for further details, let us introduce 
the \emph{character} associated to an element of $N\equiv C^\infty(\bS^2)$ as a
group homomorphism $\chi:N\to U(1)$. Since $N$ can be endowed with a
nuclear topology (see theorem 2.1 in \cite{Dappiaggi2}) it can be seen as
an element of the Gelfand triplet $N\subset L^2(\bS^2)\subset N^*$ where
$N^*$ is the set of real continuous linear functionals on $N$ (with the
induced topology). Hence, as shown in proposition 3.6 in
\cite{Dappiaggi}, for any character $\chi$ it exists a distribution
$\beta\in N^*$ such that 
\beq\label{chara}
\chi(\alpha)=e^{i(\beta,\alpha)},
\eeq 
where $(,)$ stands for the pairing between $N^*$ and $N$. 

Such a result 
can be applied also to the translational subgroup of the Poincar\`e
group on $\Im^+$ provided either that one exploits the inclusion 
$T^4\subset C^\infty(\bS^2)$ previously discussed either that the dual
space of $T^4$ - namely $\left(T^4\right)^*$ is characterised in the
following way \cite{Mc4}: if we construct the \emph{annihilator} of $T^4$ as 
$$\left(T^4\right)^0=\left\{\beta\in N^*\;|\;(\beta,\alpha(z,\bar{z})=0,
\;\forall\alpha(z,\bar{z})\in T^4\right\},$$
$\left(T^4\right)^*$ is (isomorphic to) the quotient $\frac{N^*}{\left(T^
4\right)^0}$.

Still referring to \cite{Dappiaggi}, the Wigner-Mackey approach for the 
BMS group introduces the intrinsic \emph{covariant} scalar field on null
infinity as a map $\psi:N^*\to\mch$ which transforms under the unitary
representation $D$ of $SO(3,1)\ltimes C^\infty(\bS^2)$ as
$$\left[D(\Lambda,\alpha(z,\bar{z}))\widetilde\varphi\right](\beta)=\chi
_\beta(\alpha)\widetilde\varphi(\Lambda^{-1}\beta),\quad \forall\; 
(\Lambda,\alpha(z,\bar{z}))\in SO(3,1)\ltimes C^\infty(\bS^2)$$
where $\chi_\beta$ is a character. 

Whenever the bulk spacetime is the Minkowski background and hence we
deal with the $\mathcal{R}$ subgroup of the BMS, the above
expression translates in 
\begin{equation}\label{covP}
\left\{\begin{array}{l}
\widetilde\varphi:\left(T^4\right)^*\to\bR\\
\left[D(\Lambda,\alpha(z,\bar{z}))\widetilde\varphi\right](\beta)=\chi_
\beta(\alpha)\widetilde\varphi(\Lambda^{-1}\beta) \quad \forall\;
(\Lambda,\alpha(z,\bar{z}))\in\mathcal{R}
\end{array}\right.,
\end{equation}
where now $\beta$ must be thought both as a distribution and as a
representative for an equivalence class in the coset $\frac{N^*}{\left(
T^4\right)^0}$.\\

\remark\label{coord}{It is important to point out that, in the above 
discussion, the real difference between a real scalar field on Minkowski
background and on null infinity is due to the action of the 
representation or more properly of the $U(1)$ phase factor. 

To be more precise proposition 3.2 in \cite{Dappiaggi2} grants us that, 
being $T^4$ a subspace of a locally convex topological linear space - 
namely $C^\infty(\bS^2)$, the coset $\frac{N^*}{\left(T^4
\right)^0}$ is a 4-dimensional space. Thus, if we introduce the set of
dual spherical harmonics $Y^*_{lm}$ with $l=0,1$, $m=-l,...,l$
defined as $(Y^*_{l^\prime m^\prime},Y_{lm}(z,\bar{z}))=\delta_{ll^\prime
}\delta_{mm^\prime}$, then any $\beta\in\left(T^4\right)^*$ can be
decomposed as
$$\beta=\sum\limits_{l=0}^1\sum\limits_{m=-l}^l\beta_{lm}Y^*_{lm}.$$
Hence we can extract from each $\beta$ the four-vector
\beq\label{4v}
\beta^\mu=-\sqrt{\frac{3}{4\pi}}\left(\beta_{00},\beta_{1-1},\beta_{10},
\beta_{11}\right).
\eeq 
Moreover we define the action of $\Lambda\in SO(3,1)$ on a generic
distribution $\beta\in N^*$ as
\beq\label{action}
\left(\Lambda\beta,\alpha(z,\bar{z})\right)=\left(\beta,\Lambda^{-1}
\alpha(z,\bar{z})\right)\quad\forall\alpha(z,\bar{z})\in
C^\infty(\bS^2),
\eeq 
being the action of $\Lambda$ on $\alpha(z,\bar{z})$ the one defined in
\eqref{BMS1} and \eqref{BMS2}.
A direct  inspection of proposition \ref{transl} and of the isomorphism 
between $\frac{N^*}{\left(T^4\right)^0}$ and $\left(T^4\right)^*$ shows 
that $\beta^\mu$ transforms as a covector and the quantity 
\beq\label{mass}
m^2=\eta_{\mu\nu}\beta^\mu\beta^\nu
\eeq
is $SO(3,1)$ invariant. Furthermore $m^2$ is also a Casimir for the
unitary and irreducible representation of the BMS group and hence also
for the $\mathcal{R}$ subgroup. Hence this shows that \eqref{covP}
differs from the counterpart in Minkowski background only in the
character.
}\\

The covariant scalar field \eqref{covP} does not transform under
an irreducible representation of the $\mathcal{R}$ group and, hence, in
a physical language it represents only a kinematically allowed
configuration.

On the opposite, if we look for a genuine free field, $\widetilde\varphi$ 
should transform under a unitary and irreducible representation; to 
overcome such a discrepancy we can still exploit Wigner-Mackey theory 
which calls for imposing a further constraint to \eqref{covP}. From a 
more common perspective in classical field theory this operation amounts 
to impose on $\widetilde\varphi$ the equations of motion written in the 
momenta representation; for the above scalar field it reads 
\cite{Dappiaggi}:
\beq\label{eqm}
\left[\eta^{\mu\nu}\beta_\mu \beta_\nu-m^2\right]\widetilde\varphi[\beta
]=0,
\eeq
where $\beta_\mu$ is the four vector as in \eqref{4v}. \\

\noindent Two comments on \eqref{eqm} are in due course:
\begin{enumerate}
\item the equation under analysis could be recast in the more
appropriate language of white noise calculus. In the general framework
of BMS free field theory $\widetilde\varphi[\beta]$ is a functional over 
a distribution space which is square integrable with respect to a 
suitable Gaussian measure $\mu$. Hence \eqref{eqm} should be recast in 
this scenario in terms of (multiplication) operators acting on 
$L^2(N^*,d\mu)$ and such analysis has been carried out in 
\cite{Dappiaggi2}. In this paper we can avoid such techniques exploiting 
the identification of $\left(T^4\right)^*$ with $\bR^4$ which grants us 
that \eqref{eqm} acquires the standard meaning {\it i.e.} the support of 
$\widetilde\varphi[\beta]$ is localised over the mass hyperboloid if 
$m^2\neq 0$ and over the light cone if $m^2=0$. Most importantly the 
function $\widetilde\varphi$ corresponds to an element in $L^2(\mC)$.
\item the equations \eqref{covP} and \eqref{eqm} are equivalent to a 
function transforming under a unitary and irreducible representation of 
the Poincar\'e group induced from the $SO(3)$ or from the $SO(2)\ltimes 
T^2$ little groups depending if $m\neq 0$ or $m=0$. At the same time a 
direct inspection of the analysis of chapter 3 in \cite{Dappiaggi} 
immediately shows that the representation in \eqref{covP} is nothing but 
the scalar BMS representation restricted to the $\mathcal{R}$ subgroup.
\end{enumerate}

Before concluding our analysis we still need the last ingredient which
relates the two above constructions of a massless scalar field theory on 
$\Im^+$.\\ 

\teorema\label{double}{\em A field $\Psi$ on $\Im^+$ constructed as in
proposition \ref{proj} corresponds to a $\mathcal{R}$ field \eqref{covP} 
which satisfies \eqref{eqm} with $m=0$. Hence the representation 
$A^{(1)}(g)$ is also irreducible on $L^2(\bR\times \bS^2,EdEd\bS^2)$.}

\begin{proof}
We provide here a much shorter proof than that of theorem 3.35 in
\cite{Dappiaggi}. Let us recall that, according to lemma \ref{un1}
$\Psi$ satisfies \eqref{BMS}.

Furthermore, following the characterisation of a light cone imbedded in 
$\bR^4$ as discussed at the beginning of section \ref{FHLC} and 
identifying $E$ with $r\doteq|\vec{p}|$ we end up with 
$\widehat\Psi\in L^2(\mC)$ being $\widehat\Psi$ the Fourier transform of
$\Psi$ constructed as in proposition \ref{proj}. According to theorem 1 
in \cite{Strichartz3}, $\widehat\Psi$ can be read on its own as the 
restriction on $\mC$ of the Fourier transform for a function 
$\widetilde\Psi$ satisfying D'Alambert wave equation and, hence, lying 
in $L^4(\bR^4,d^4x)$. The Fourier transform for $\widehat{\widetilde\Psi}
\in\mS^\prime(\bR^4)$  satisfies the constraint $\eta^{\mu\nu}p_\mu p_
\nu\widehat{\widetilde\Psi}=0$ and the Poincar\'e group $\mathcal{R}$ 
still acts as $A^{(1)}(g)$.

To conclude the demonstration, let us now consider \eqref{covP} which
satisfies \eqref{eqm} with $m=0$.
Exploiting the identification between the distribution $\beta$ and the
covector $p_\mu$, a direct inspection shows that the scalar $\mathcal{R}$ 
representation acts on \eqref{covP} as the representation $A^{(1)}(g)$.
Thus each $\Psi$ constructed in proposition \ref{proj} has been mapped
into a massless $\mathcal{R}$ scalar free field. Irreducibility of
$A^{(1)}(g)$ is now a consequence of Mackey construction which grants us
that the scalar $\mathcal{R}$ (and, thus, the $A^{(1)}(g)$) 
representation induced from the scalar $E(2)$ representation is 
irreducible.
\end{proof}

\noindent We have now all ingredients to conclude our analysis on the projection of
a massive bulk scalar field:\\

\teorema\label{2KG}{\em Let us consider any norm-finite solution $\phi$ 
of \eqref{KG} with the associated triplet $(\psi, U^\prime, \widetilde T
)$. The latter projects to a triple $(\Psi, U^\prime, \widetilde T)$ on
future null infinity which identifies two Poincar\'e invariant free 
scalar field constructed \`a la Wigner-Mackey and solving \eqref{eqm} 
with the same mass value as $\phi$.}

\begin{proof}
According to the hypothesis of the theorem we can associate to $\phi$
the triplet $(\psi, U^\prime, \widetilde T)$ where $\psi$ can be written
as \eqref{massless2}. We can now exploit proposition \ref{proj} to
project $\psi$ in a square integrable function $\Psi$ over $\Im^+$:
$\Psi=\rho(\widetilde\psi)$ where $\widetilde\psi\doteq\Omega^{-1}\psi$. 
Hence, being $U^\prime$
and $\widetilde T$ respectively a representation and an intertwiner
thus independent from coordinates, we construct on null infinity the
triplet $(\Psi, U^\prime, \widetilde T)$. The representation
$U^\prime(\Lambda)$ is the quasi-regular representation of the Lorentz
group and it unambiguously induces (or it is the restriction of) the
scalar $\mathcal{R}$ representation which acts on $\Psi$ as the
representation $A^{(1)}(g)$ from \eqref{BMS}. We can now exploit theorem
\ref{double} according to which the pair $(\Psi, A^{(1)}(g))$
corresponds to one $\mathcal{R}$ invariant field
\eqref{covP} which satisfies \eqref{eqm} with $m=0$. Hence $(\Psi,
A^{(1)}(g))$ can be traded with $(\widetilde\varphi,
D(\Lambda,\alpha(z,\bar{z}))$ where
$\left(\Lambda,\alpha(z,\bar{z})\right)\in\mathcal{R}$ and $D$ is the scalar
representation in \eqref{covP}. 

Still the induction-reduction theorem for group representation (chapter 18 in
\cite{Group}) grants us that the restriction of $D$ to $SO(3,1)$ is
exactly $U^\prime(g)$ and that the quasi-regular representation
unambiguously induces the scalar $\mathcal{R}$ representation. Hence we
have mapped the original triplet $(\Psi, U^\prime, \widetilde T)$ in 
$(\widetilde\varphi, U^\prime, \widetilde T)$. The circle has been
almost closed and our last step consists of exploiting the same reasoning 
as in the proof of theorem \ref{double} {\it i.e.} we can read 
$\widetilde\varphi$ as a solution from the massless wave equation 
constructed out of an element of  $L^2(\mC)$ - say $\hat{\widetilde
\varphi}|_\mC$. Hence we can now exploit our last ingredient namely the 
intertwiner $\widetilde T: L^2(\mC)\to L^2(\bH_m)\oplus L^2(\bH_m)$ 
{\it i.e.} $\widetilde T(
\hat{\widetilde\varphi}|_\mC)=(f,g)$. Accordingly both $f$ and
$g$ lies in $L^2(\bH_m)$ and the Lorentz group acts as $\left[U^\prime(
\Lambda)f\right](p^\mu)=f(\Lambda^{-1}p^\mu)$ for all $\Lambda\in 
SO(3,1)$. Still exploiting theorem 1 in \cite{Strichartz3}, we can 
interpret $f$ (or $g)$ as the restriction on the mass hyperboloid $\bH_m$ 
of a function - say $\widetilde\varphi_f$ or $\widetilde\varphi_g$ - 
whose Fourier transform satisfies the Klein-Gordon equation of motion 
with mass $m$ and it lies in $L^p(\bR^4, d^4x)$ with $\frac{10}{3}\leq p
\leq 4$. If we now take into account that the original field 
$\widetilde\varphi$ is an intrinsic $\mathcal{R}$ free field, we are
entitled to switch from $p^\mu$ to the variables $\beta^\mu$. To
conclude we can exploit remark \ref{coord} according to which a covector
$\beta^\mu$ transforming under the standard $SO(3,1)$ action corresponds
to a distribution $\beta\in\left(T^4\right)^*\subset N^*$ on which
$\Lambda\in SO(3,1)$ acts according to \eqref{action}. Eventually still 
the induction theorem allow us to construct from $U^\prime(\Lambda)$ the
scalar $\mathcal{R}$ representation $D$. Hence both $(\widetilde
\varphi_f, U^\prime(g))$ and $(\widetilde\varphi_g, U^\prime(g))$
correspond unambiguously to a $\mathcal{R}$ massive scalar field as in 
\eqref{covP} with support on the mass hyperboloid {\it i.e.} with the 
same value for $m^2$ as the original Minkowski field $\phi$.\\
\end{proof}

\remark{The projection of a bulk massive scalar field into two boundary
massive scalar fields is a natural byproduct of the intertwining
operator. In the projection of $f\in L^2(\bH_m)$ to a function 
over the light cone, we could imbed $f$ into the element $(f,f)$ of the 
diagonal subgroup of $L^2(\bH_m)\oplus L^2(\bH_m)$; on the opposite on
the boundary we perform the inverse operation mapping a square
integrable function over the light cone into 
$L^2(\bH_m)\oplus L^2(\bH_m)$. Hence there is no guarantee that the
intertwiner identifies an element of the diagonal subgroup and we are
forced to take into account two massive fields instead of a single one.}

\section{Conclusions}
In this paper we have shown that, exploiting Strichartz harmonic
analysis on hyperboloids, it is possible to project the information of
a norm finite massive real scalar field $\phi$ in Minkowski spacetime 
into null a triplet of data on null infinity: $(\Psi, U^\prime, 
\widetilde T)$ where $\Psi$ is the projection on $\Im^+$ out of trace 
operator of a solution for the D'Alambert wave equation, $U^\prime$ is 
the $SO(3,1)$ quasi-regular representation whereas $\widetilde T$ is a 
unitary intertwiner from $L^2(\mC)$ to $L^2(\bH_m)\oplus L^2(\bH_m)$. 

The result we achieve has a twofold advantage. From one side it is
coherent with Helfer result which states that the space of section for
any vector bundle over null infinity carries only the massless
representation for the homogeneous action of the Poincar\'e group. As a
matter of fact $\Psi$ can be ultimately interpreted as a free field on
the conformal boundary with $m=0$. From the other side we can recover the
original interpretation of massive fields exploiting the action of
$\widetilde T$ and, as shown in theorem \ref{2KG}, the original single
field $\phi$ corresponds to two separate massive free fields in the 
$\mathcal{R}$ invariant theory constructed \`a la Wigner-Mackey.

Although we believe the result is rather appealing opening a wide range
of possible applications, it is fair to admit that it is to a certain
extent not sharp. As a matter of fact, in the whole construction we
performed three arbitrary choices: the first, already discussed, refers
to the imbedding of an element $f\in L^2(\bH_m)$ (the restriction on the
mass hyperboloid of the Fourier transform of $\phi$) into the diagonal
subgroup of $L^2(\bH_m)\oplus L^2(\bH_m)$. 

The second and the strongest between the performed choices arises in the 
projection to $\Im^+$; the general solution $\widetilde\psi$ of the 
D'Alambert wave equation we constructed lies in $L^4(M^4,\sqrt{|g|}d^4x)$ 
but, in order to apply trace theorems, we needed to consider at least 
Sobolev spaces of first order. This restricts the range of validity of 
our results and it will be interesting to eliminate such constraint from 
our analysis.

The third and less pernicious of the choices lies in the construction of
the above mentioned trace operator. As a matter of fact we embedded
Minkowski spacetime into an open region of Einstein static universe.
Hence this amount to select a preferred gauge factor $\omega$ according
to the definitions of section $\ref{AFS}$ contrary to the projection
operator introduced in $\cite{Dappiaggi, Mason}$ which provides a
smooth function over $\Im^+$ for any possible choice of $\omega$.
Nonetheless we feel that, fixing $\omega$ in our analysis, does not lead 
to a loss of generality since we can ultimately interpret our results in
terms of a general field theory constructed over $\Im^+$ without the
need for a choice for the gauge factor.     

To conclude we wish to discuss possible applications of our results. Our
main target is an holographic interpretation of bulk field theory along
the lines of \cite{Dappiaggi} and the previous section was written with
this goal in mind. As a matter of fact we have proved that, at least in
Minkowski background, it is possible to project each solution of a
massive Klein-Gordon equation of motion into a suitable counterpart at
null infinity. Such bulk-to-boundary interplay does not represent the
only possible application of our analysis and we envisage that our
results could be possibly exploited for other research fields such as,
to quote an example, conformal scattering problems.

Nonetheless we believe that the most interesting perspective consists of 
the development of a similar result for generic globally hyperbolic and 
asymptotically flat spacetime. Already at a first reading of this 
manuscript, one can realize that the extensive use of tools proper of 
harmonic analysis forbids to mimic our procedure in a more
generic scenario\footnote{It is possible to wonder if our construction
extends to higher dimensional flat backgrounds and, since both the tools 
proper of harmonic analysis and Strichartz estimates can be generalised
to any dimension, there is no apparent obstruction. On the opposite the
question if one can derive a similar result for curved backgrounds is
less adequate since conformal completion techniques \`a la Penrose may
run into serious difficulties \cite{Hollands}.}. Nonetheless we feel that 
finding a way to project the information for a massive scalar field on 
null infinity in Minkowski background is an encouraging starting point to 
deal with the same problem in more complicated frameworks. A positive 
conclusion of such a research project would also open the way to develop 
in a generic background (and not only in Minkowski), with tools proper of 
the algebraic quantisation of field theory, a correspondence between the
quantised bulk massive scalar theory and the bulk counterpart. We aim to
address such an issue in future works.\\   

\begin{center}
\Large{\bf Acknowledgments}
\end{center}

\vskip .2cm

\noindent The author is grateful to V. Moretti and N. Pinamonti for 
fruitful discussions at early stages of this research project and to O. 
Maj for several usefull discussions on hyperbolic partial differential 
equations. This work has been partly supported by GNFM-INdAM (Istituto 
Nazionale di Alta Matematica) under the project \emph{``Olografia e 
spazitempi asintoticamente piatti: un approccio rigoroso''}.

\end{document}